# Design of a High Intensity Neutron Source for Neutron-Induced Fission Yield Studies

M. Lantz[1], D. Gorelov[2], A. Jokinen[2], V.S. Kolhinen[2], A. Mattera[1], H. Penttilä[2], S. Pomp[1], V. Rakopoulos[1], S. Rinta-Antila[2], A. Solders[1]

1) Physics and Astronomy, Applied Nuclear Physics division, Uppsala University, Box 516, SE-75120 Uppsala, Sweden
2) Department of Physics, P.O. Box 35 (YFL), FI-40014 University of Jyväskylä, Finland

Corresponding author: Mattias.Lantz@physics.uu.se

**Abstract**. The upgraded IGISOL facility with JYFLTRAP, at the accelerator laboratory of the University of Jyväskylä, has been supplied with a new cyclotron which will provide protons of the order of 100 µA with up to 30 MeV energy, or deuterons with half the energy and intensity. This makes it an ideal place for measurements of neutron-induced fission products from various actinides, in view of proposed future nuclear fuel cycles. The groups at Uppsala University and University of Jyväskylä are working on the design of a neutron converter that will be used as neutron source in fission yield studies. The design is based on simulations with Monte Carlo codes and a benchmark measurement that was recently performed at The Svedberg Laboratory in Uppsala. In order to obtain a competitive count rate the fission targets will be placed very close to the neutron converter. The goal is to have a flexible design that will enable the use of neutron fields with different energy distributions. In the present paper, some considerations for the design of the neutron converter will be discussed, together with different scenarios for which fission targets and neutron energies to focus on.

## 1. Introduction

The Applied Nuclear Physics division at Uppsala University and the IGISOL group at the Department of Physics at University of Jyväskylä are collaborating, with the purpose of measuring neutron-induced independent fission yields of different actinides of relevance for partitioning and transmutation of spent fuel and for other aspects where information on nuclear fuel inventories are important. The project will use the upgraded IGISOL-JYFLTRAP facility at the accelerator laboratory of the University of Jyväskylä. The Jyväskylä group is on the forefront when it comes to accurate measurements of reaction products from nuclear interactions involving short lived nuclei. With the Ion Guide Isotope Separator On-Line (IGISOL) technique high yields of reaction products are selected and then accurately determined through mass measurement in the JYFLTRAP Penning trap. This method has proven to be very useful for the determination of independent fission yields, and experiments have been performed with 20-50 MeV protons on $^{232}$Th and $^{238}$U, and with 25 MeV deuterons on $^{238}$U [1].

In order to measure neutron-induced independent fission yields some sort of neutron source is needed. The IGISOL facility was recently moved to a new experimental hall where it is supplied with a new cyclotron which provides proton beams of the order of 100 µA with 18-30 MeV energy, or deuterons of half that energy and intensity [2]. Therefore a neutron converter is designed, utilizing (p,xn) or (d,xn) reactions. Different options have been investigated, and our approach is to use protons on a water cooled Beryllium plate through the Be(p,xn) reaction. The design is based on simulations with Monte Carlo codes such as FLUKA [3, 4] and MCNPX [5], and deterministic codes such as COMSOL Multiphysics [6]. There will always be uncertainties in the predictions given by the Monte Carlo codes. Therefore a benchmark measurement was performed in June 2012 at The Svedberg Laboratory (TSL) in Uppsala, funded through the ERINDA EU framework programme [7].

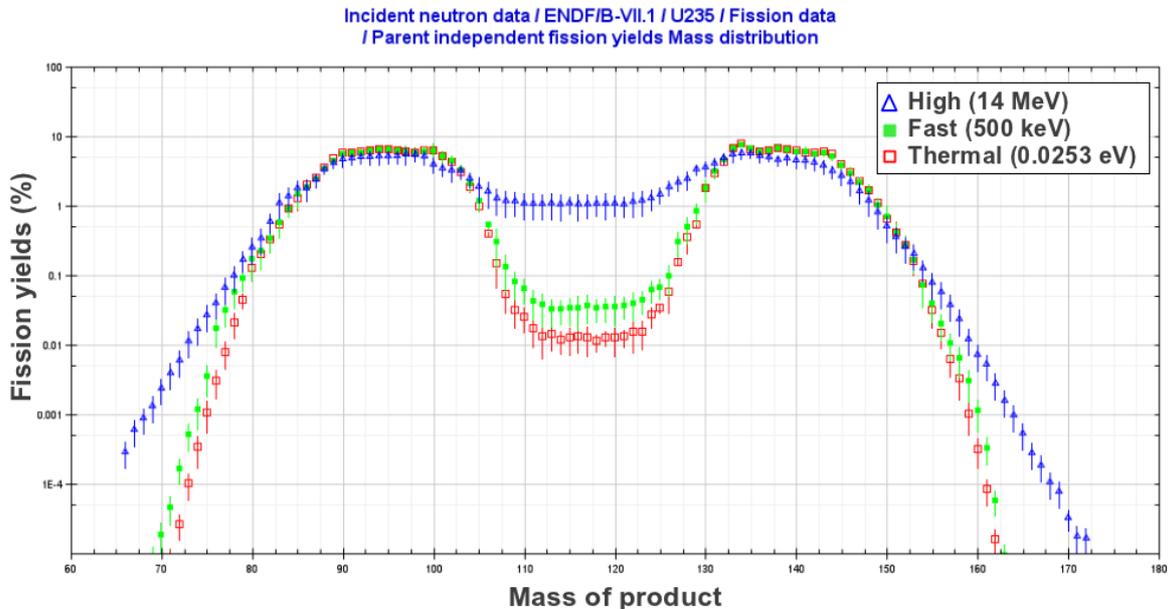

FIG. 1. Fission yield mass distributions for thermal (0.0253 eV), fast (500 keV) and high energy (14 MeV) neutrons on $^{235}$U from the ENDF/B-VII.1 evaluation. The plot is produced with JANIS [11].

In order to obtain competitive count rates the fission targets will be placed very close to the neutron converter [8]. The goal is to have a flexible design that will enable the use of neutron fields with different energy distributions. Different fields can be obtained by varying the proton energy, inserting moderator materials, using deuterons instead of protons, or by varying the converter material and thickness. Challenges for the design include the need for cooling, how to avoid too much activation of the converter material and an intense background of low energy neutrons and high energy photons.

## 2. Independent Fission Yields

Accurate knowledge about fission yields distributions is of importance for a better theoretical understanding of the fission process itself. Besides theoretical development there are also a number of applications related to nuclear power generation where better knowledge is beneficial, for instance [9]:
- information about the composition of the resulting spent fuel (spent fuel repositories, partitioning and transmutation issues, Gen-IV reactor scenarios),
- various safety measures (decay heat, fission gas production, criticality, dosimetry, safeguards, delayed neutrons),
- information about neutron poisoning (significant discrepancies between different evaluations have been identified, especially for $^{135}$Xe, $^{149}$Sm and $^{157}$Gd [10]), and
- improvement of burn-up predictions.

The well known shape of the fission yield mass distribution with two peaks around mass number A = 95 and A = 135 is characteristic for the thermal neutron field in a light water reactor (LWR) with $^{235}$U. But as shown in FIG. 1 the mass distribution varies with neutron energy, and it also depends on the initial actinide. Therefore systematic measurements at different neutron energies and for different fission target nuclides are of importance. Furthermore, the mass distribution seen in FIG. 1 reflects the situation for many experiments where only the masses of the fission products are determined, though usually in combination with other observables such as kinetic energy [12]. Therefore it is valuable to obtain independent fission yields as complementary information to other data.

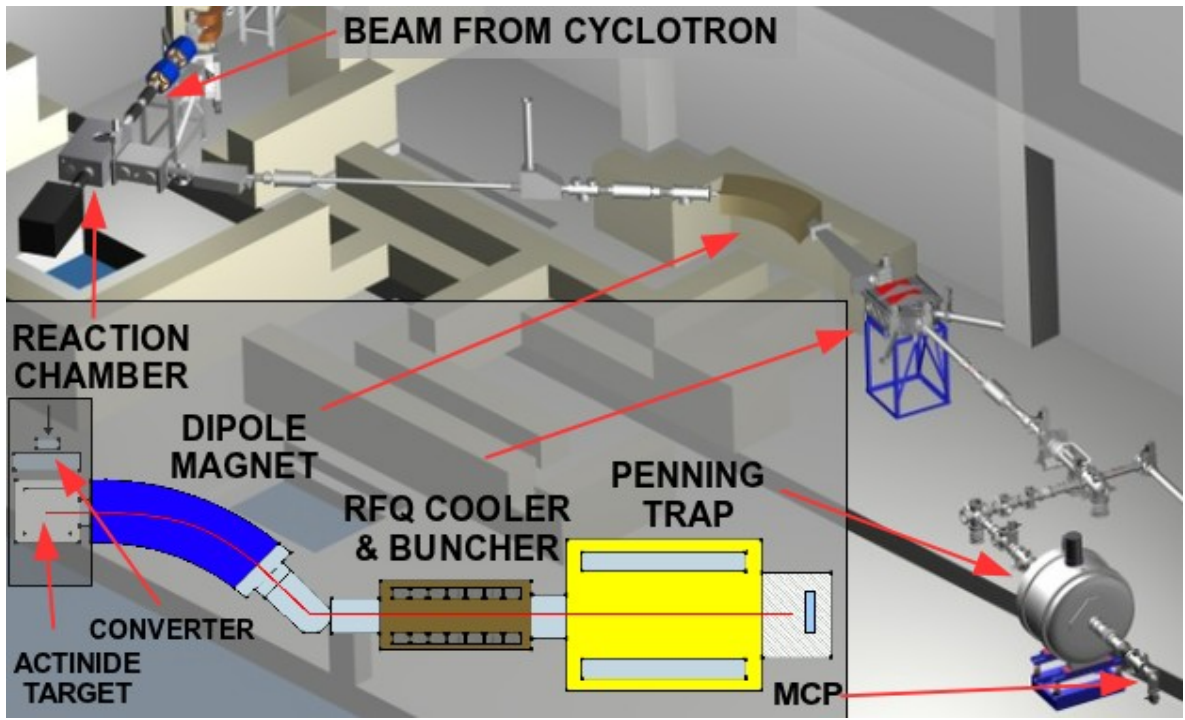

*FIG. 2. Layout of the IGISOL-JYFLTRAP facility (schematic view in the inset). A particle beam enters from the top left impinging on a neutron converter, resulting in a neutron field causing fission in an actinide target. The fission products are accelerated and transported through a dipole magnet for mass selection, followed by an RFQ cooler and buncher for improvement of the beam properties. The cooled ions are injected into the Penning trap for accurate mass determination and finally the count rate is determined by sending the ions to an MCP detector. CAD drawing by Tommi Eronen.*

It should be mentioned that independent fission yields are defined as the percentage of atoms of a specific nuclide produced directly in fission reactions, i.e., before radioactive decay has occurred [9].

## 3. The experimental technique

The description of the experimental technique applies both for charged particle beams and neutron beams, and the principles are the same irrespective of if it is fission products or other kinds of reaction products that are measured. The principle layout of the IGISOL-JYFLTRAP facility is shown in FIG. 2 for the case of using a neutron converter and moderator in order to measure independent fission yields with neutrons down to thermal energies. The considerations for a neutron converter are discussed in Section 4.

### 3.1. Particle beams

The recent move of the IGISOL facility to a new experimental area means a general upgrade of the facility. An important addition is a new MCC30/15 cyclotron [2] providing protons in the energy range 18-30 MeV and deuterons of 9-15 MeV. For protons a beam current of 100 µA or more is expected, making it possible to consider high intensity neutron beams. The new cyclotron will commence operation during 2013. Together with the existing K-130 cyclotron more than 4000 hours of beam time can be provided annually for the IGISOL facility.

## 3.2. Reaction (fission) chamber and ion guide

The reaction chamber contains the beam target from which reaction products or fission products are ejected. The target has to be thin enough to allow the fission products to escape from it. Helium or other noble gases are used as buffer gas that flows through the reaction chamber and slows down all fission products with energies less than 2 MeV. The gas pressure in the chamber is about 200 mbar. Fission products with higher energies are lost as they will hit the chamber walls. The fission products are initially completely ionized, but rapidly change their charge state through interactions with the buffer gas. Due to the high ionization potential of noble gases, a large fraction of the products retain a $1^+$ charge state. The gas flow transports the fission products out of the chamber into an ion guide that centers them on the axis and accelerates them in steps to about 30 kV [13].

## 3.3. Mass selection, beam refinement, and mass determination

The accelerated ions are sent through a 55° dipole magnet for a first mass selection, with mass resolving power up to 500. At this stage the ions have a relatively large transverse emittance and energy spread. Therefore they are refined in a gas-filled Radio Frequency Quadrupole (RFQ) buncher and cooler before being injected to the JYFLTRAP two stage Penning trap. The Penning trap can be used to determine the mass of an ion by finding its cyclotron frequency in a strong magnetic field. The cyclotron frequency of the oscillating ion can be probed by applying an alternating quadrupole potential to a set of segmented ring electrodes and the mass can be determined through the relation $f_c=(1/2\pi)(Bq/m)$. Here $f_c$ is the cyclotron frequency of the ion with charge q and mass m, oscillating in an external magnetic field B.

The trap can also be used as a high resolution mass filter. By first subjecting the ions to a dipole oscillating field, followed by a mass selective quadrupole field combined with a buffer gas, only those ions for which the frequency of the applied field matches the cyclotron frequency are selected. Finally the selected ions are ejected from the trap through a narrow aperture and are detected at a Multichannel Plate Detector (MCP) where the count rate is measured as a function of the quadrupole frequency. This method selects ions with a mass resolving power of up to $10^5$ which means that it is possible, through peak fitting, to resolve metastable states that are 0.5 MeV apart [14].

The total time from fission product emission in the reaction chamber to detection in the MCP is a few hundred milliseconds, enabling the determination of the independent fission yields for a large selection of nuclides.

## 3.4. Potential difficulties with the ion guide technique

The experimental method has some potential limitations that need to be considered [1]:
- The low stopping power of the helium gas only stops 1-10% of all fission products. This could potentially lead to a bias in which fission products that are being studied.
- To a first approximation the ion guide technique is insensitive to chemical properties as ions of any element can be produced. But some elements rapidly form oxides, and there are several elements that tend to be extracted as $2^+$ ions.
- It may be relevant to consider whether there is some sort of mass dependence in how well different fission products are transported by the ion guide.

- For fast neutrons there may be non-isotropic effects on the spatial distribution of fission products due to the fact that the incident particle brings high momentum into the fissioning system.

A dedicated investigation was performed by comparing the isotopic yields of Rb and Cs isotopes obtained in proton-induced fission of $^{238}$U [13, 15] with high quality data from a different experimental method [16]. The results were found to be in good agreement. Further cross checks have been performed by comparing results with some experiments performed at Tohoku University [17]. The Tohoku experiments use a similar ion guide technique, but the different geometry compared with IGISOL allows useful inter-comparisons of several of the concerns, usually with reasonable agreement [1].

Also the Penning traps have chemical effects to consider. Several cross checks, including calibration with alpha recoil sources or fission sources placed inside the ion guide, have been performed or have been suggested. Other concerns are corrections for decaying isomers, the accumulation of decay products in the trap, and time dependent variations of the count rate. An important remedy for the latter uncertainty is that for all measurements there are data taken during the same run for known reference masses, enabling data renormalization [1].

## 4. The neutron source

Although most fission yield experiments at IGISOL were performed with protons, two attempts have been made with neutrons, showing that the ion guide technique is feasible also for such measurements [18, 19]. In those tests the $^{12}$C(d,xn) and $^{13}$C(p,xn) reactions were used, and the incident beam currents were a few μA. In the present project higher neutron yields are sought through proton- or deuteron-induced reactions on other target materials, and with the new MCC30/15 cyclotron that provides higher beam currents.

### 4.1. Design considerations

For the introduction of a neutron source to the IGISOL facility there are several issues to consider:
- Neutron yield: In order to be competitive in comparison with other experimental facilities, in studies of nuclides far from the stability line, the neutron converter should be able to deliver about $10^{12}$ fast neutrons ($E_n > 1$ MeV) on a $^{238}$U target.
- Neutron energy spectra: For studies of independent fission yields of relevance for nuclear power applications the incident neutrons should have an energy distribution resembling those in light water reactors (LWR) or fast reactors. Mono-energetic neutrons are also considered.
- Cooling issues: With 30 MeV protons of 100 μA beam intensity or more, at least 3 kW of heat is deposited into a very small volume of the neutron converter. Therefore sufficient cooling of the converter has to be provided, and the number of suitable converter materials becomes limited.
- Activation and structural integrity: The neutron converter will become activated through the reactions with protons or deuterons, and the produced neutrons will activate surrounding materials. This may reduce access to the facility after irradiation. Furthermore, the converter may suffer structural problems through hydrogen buildup if it is thick enough to fully stop protons or deuterons. For thinner targets the residual beam may activate other material.

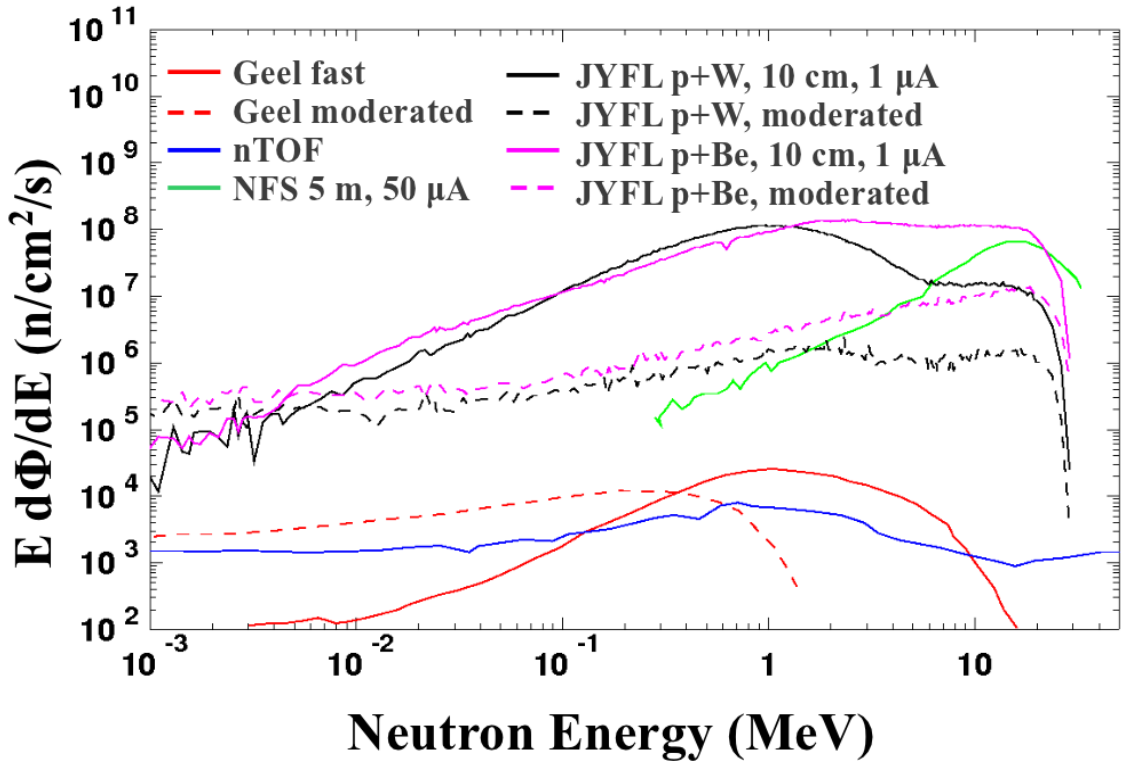

*FIG. 3. Isolethargic neutron spectra calculated for thick Be and W targets considered for the Jyväskylä laboratory (JYFL), compared with other neutron beams. It should be noted that the calculations are shown for 1 µA protons, the actual neutron flux may thus be increased by two orders of magnitude. Calculations were made with FLUKA [3, 4].*

- Flexible design: Several of the issues above can be handled by using different materials and thicknesses. This requires a design where the converter easily can be replaced with a new one, without complicated issues of breaking vacuum and risk of leakage of the cooling media. Furthermore, for toxic materials such as Be, a design is required that reduces the amount of handling and machining of the converter material.

With these issues in mind Be and W converters with different geometries have been considered [8]. Both materials have high melting points and heat transfer properties, and may be coupled mechanically to a cooling device. The requirement of high neutron intensity may be fulfilled by placing the converter very close, about 10 cm, to the fission target. Monte Carlo codes have been used in order to estimate the neutron flux for different options. FIG. 3 shows neutron spectra for Be and W converters, with and without moderator material, compared with other neutron facilities. It is noteworthy that due to the close proximity of the fission target the overall neutron yield will be much higher than for any of the other facilities. Finally, a Be converter gives many more fast neutrons than W and therefore it was decided to use it for the converter.

### 4.2. Simulations

Monte Carlo simulations have been performed both with FLUKA [3, 4] and MCNPX [5]. As seen in FIG. 4 it is relatively easy to moderate neutrons in such a way that the low energy part resembles that of an LWR, while it is difficult to obtain fast neutron spectra similar to those in fast reactors. For the fast spectra the simulations also reveal a significant discrepancy in the low energy part between the two Monte Carlo codes, the reason for this has not been determined yet.

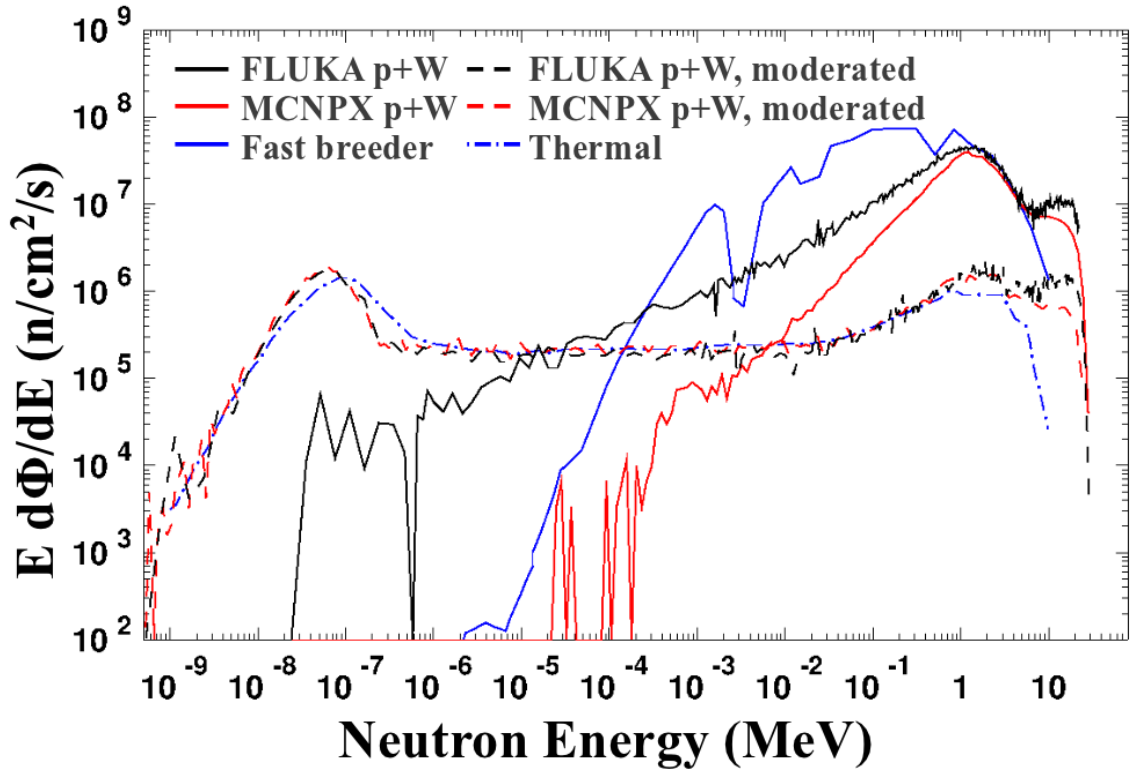

*FIG. 4. Comparison of neutron spectra from 30 MeV protons on a thick W converter, with and without 10 cm water moderator, simulated with FLUKA (black) and MCNPX (red). Typical LWR (thermal) and fast breeder spectra (blue) are shown for comparison [20]. The reactor spectra have been arbitrarily scaled for the comparison.*

Studies of the need for cooling of the converter have been performed with COMSOL Multiphysics [6]. Both for Be and W it has been possible to identify geometries thick enough to fully stop 30 MeV protons that can be sufficiently cooled by a flow of water in a closed loop.

### 4.3. Approaching a final design

Taking into account all the initial criteria, and the results from the simulations, a preliminary design has been agreed on that fulfills most of the criteria. The design is inspired by the LENS target developed at Indiana University Cyclotron Facility (IUCF), where a Be target slightly thinner than the full proton stopping length is used [21]. The target is assembled in a holder and acts as a window between the evacuated beam pipe on the upstream side, with a 0.5 cm thick layer of cooling water on the downstream side. By not stopping the protons within the target itself the cooling requirement is drastically reduced, as is the risks of structural degradation from hydrogen buildup. The reduction in neutron yield is about 5%. The main drawback is that there may be chemical effects in the cooling water due to the stopping of protons in it. In order to avoid exposure to beryllium dust, target cylinders will be bought at standard dimensions from commercial utilities and be used without any further machining. The holder uses O-rings on both sides of the plate in order to ensure an airtight assembly. Furthermore a second window, made of steel or havar, will be inserted in the beam pipe in order to reduce the effects in case of a water leak. A preliminary design is shown in FIG. 5.

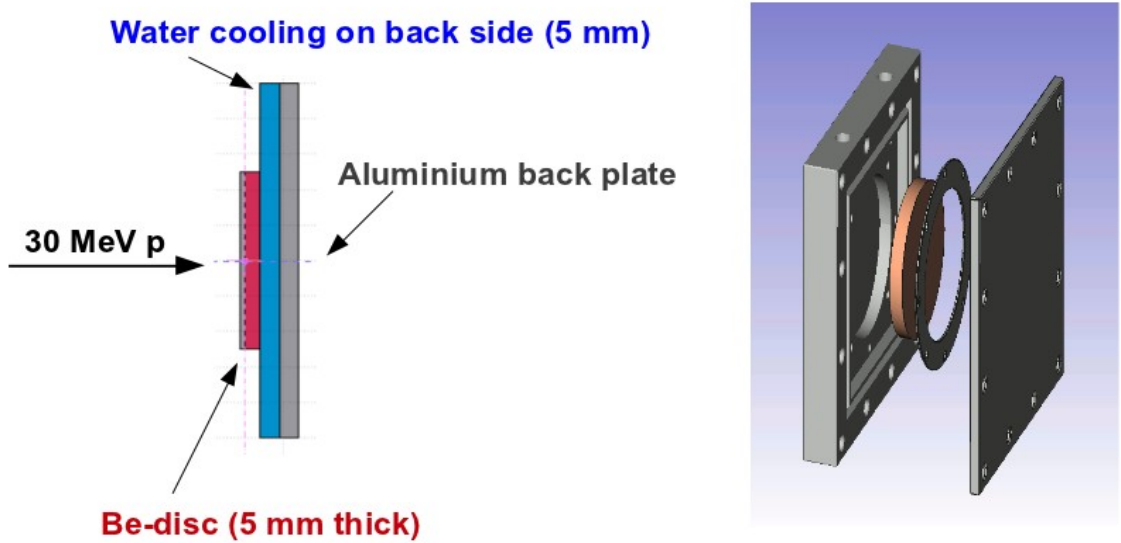

*FIG. 5. Principle design of a beryllium neutron converter for the IGISOL setup.*

### 4.3. Benchmark experiment

Due to the discrepancies in predictions between FLUKA and MCNPX it was necessary to compare the simulations with experimental data. The few experiments performed in the relevant energy range all have some question marks of relevance, and none of them cover the full energy range. Therefore it was deemed necessary to perform a reference measurement that could provide some guidance on how to interpret the simulated results. Through the EU-funded ERINDA framework programme, beam time was obtained for a measurement of the neutron energy spectra from a Be converter resembling the preliminary design. The experiment was performed in 2012 at The Svedberg Laboratory, using 37 MeV protons degraded to 30 MeV. Outgoing neutrons were measured with Bonner Sphere Spectrometers, covering the range from thermal energies up to about 20 MeV, with a $^6$Li(Eu) scintillator as thermal neutron detector [22]. In parallel a NE-213 liquid scintillator, allowing n-γ pulse shape discrimination, was used for a Time-of-Flight (ToF) measurement in the energy range 5-30 MeV, thus providing a good overlap region for intercomparisons between the two methods [7]. The ToF data were saved on two different data acquisition systems, one with pre-set thresholds, and one where the detector pulses were saved event by event for off-line analysis.

The experiment was performed with different settings, including background measurements with shadow cones and with slow neutrons moderated through polyethylene. For the ToF measurement it was also possible to vary the thickness of the Be converter [23]. It is important to stress that the background conditions at TSL will be very different from those at IGISOL due to different geometries around the converter. By having a controlled experiment to compare the simulations with it will be possible to make better predictions of the neutron spectrum that induces the fission yields measured at IGISOL.

Simulations showed that the low energy neutron background at TSL was significant. This will also be expected at IGISOL and has to be taken into account. For measurements on actinides such as $^{238}$U with a high energy threshold for fission the background is of no importance, while for actinides with high fission cross sections at thermal energies, such as $^{235}$U, the effects may be significant. Furthermore, high energy photons may induce photofission. While having relatively low cross sections, the effect should be estimated from the measurements and from simulations. The analyses from the measurements will be finalized during the spring 2013 [7, 22, 23] helping in deciding the final design of the neutron converter to be used at IGISOL.

**5. Outlook**

The first experiments at the upgraded IGISOL-JYFLTRAP facility are planned for the late spring of 2013, including a measurement of isomeric yield ratios from 25 MeV protons on $^{238}$U. Thereafter the commissioning of the neutron converter will start. The plan is to begin with neutron-induced fission from a large $^{238}$U foil with neutrons from a Be converter. Thereafter a number of parameters can be changed in order to obtain different neutron energy distributions; reduced proton energy, use of deuterons, use of thin target, and use of different converter material. For thin target measurements semi-monoenergetic neutron beams will provide reference points. None of the neutron fields will agree with those in a reactor, but from measurements with different fields the energy dependence of the independent fission yields can be obtained and used in unfolding.

Another parameter to change is the fission target. Although it is the most obvious thing to change, it may be very difficult to obtain different target actinides of interest. Besides the practical difficulty of obtaining and manufacturing targets from some fissile nuclides, there may be regulatory restrictions making transfer of the material between different countries difficult or even impossible. It may seem a bit ironic that research of relevance for the development of safer nuclear power and safer handling of used fuel may be hindered due to rules regulating the safe handling of the very same nuclides. A preliminary wish list includes the following actinides:
- Relatively easy to access and handle: $^{235}$U, $^{238}$U, $^{232}$Th
- Somewhat more difficult to obtain: $^{237}$Np, $^{239}$Pu, $^{234}$U
- Very difficult to manufacture or handle: $^{240}$Pu, $^{243}$Am

In spite of these challenges there will be plenty of experiments to be done starting with $^{238}$U, both from an applications point of view and for fundamental research in nuclear structure. It is the ambition of the groups in Jyväskylä and Uppsala to make a long term commitment to this project. For a long term dedicated effort follows some challenges with respect to securing funding and to have continuity in manpower, not the least in order to attract talented students. It should be emphasized that the EU-funded programmes for transnational laboratory access are increasingly important in order to enable these kinds of projects, and for providing people with relevant expertise both for research and industry.

**6. Summary**

The IGISOL-JYFLTRAP experimental programme at University of Jyväskylä has been successful in developing the method of combining the ion guide technique with proton-induced independent fission yields, enabling high accuracy mass determinations of the fission products. The new high intensity cyclotron makes it feasible to initiate measurement campaigns of neutron-induced independent fission yields. A beryllium neutron converter is planned for the facility, and a flexible design enables measurements with many different neutron fields, providing nuclear data of importance for future nuclear power systems and for the safe handling of used nuclear fuel. In order to quantify the neutron energy distributions at IGISOL a reference measurement has been performed at The Svedberg Laboratory with an experimental setup that resembles the preliminary design. The reference measurement will help with interpretations of predictions from Monte Carlo codes and may help in further benchmarks of those codes. The described project is in its initial phase, but the involved groups view it as a long term commitment where nuclear data for society and fundamental research go hand in hand.

## 7. Acknowledgements

The work was supported by the European Commission within the Seventh Framework Programme through Fission-2010-ERINDA (project no. 269499), by the Swedish Radiation Safety Authority (SSM), and by the Swedish Nuclear Fuel and Waste Management Co. (SKB).